\documentclass[12pt,a4paper]{article}
\usepackage{amssymb}

\input{tcilatex}

\begin{document}

\title{Possible cosmological implications of a time varying fine structure constant.}
\author{Marcelo.S.Berman$^{(1)}$ \and and Luis.A.Trevisan$^{(2)}$ \\
(1) Tecpar-Instituto de Tecnologia do Paran\'{a}.\\
Grupo de Projetos Especiais.\\
R.Prof Algacyr M. Mader 3775-CIC-CEP 81350-010\\
Curitiba-PR-Brazil\\
Email: marsambe@tecpar.br\\
(2) Universidade Estadual de Ponta Grossa,\\
Demat, CEP 84010-330, Ponta Grossa,Pr,\\
Brazil \ email: latrevis@uepg.br}
\maketitle

\begin{abstract}
Webb et al. [2] experimental results on the fine strucutre \ ``constant ''
variation with the age of the Universe, is here analized. By using \ the
experimental data on the fine-structure ``constant '''s variation with the
age of the Universe, and Dirac's LNH (Large Number Hypothesis), we find how
should vary the total number of nucleons in the Universe, the speed of
light, Newton's gravitational ``constant '' and the energy density, and we
make an estimate on the deceleration parameter, finding that the Universe
would be accelerating, just as Supernovae observations have concluded.

PACS 98.80 Hw
\end{abstract}

\newpage

\begin{center}
POSSIBLE COSMOLOGICAL IMPLICATIONS OF A TIME VARYING FINE STRUCTURE CONSTANT.

MARCELO S. \ BERMAN and LUIS A. TREVISAN
\end{center}

\noindent

In the two landmarking letters, Webb et al. [1] and Webb et al.[2] have
provided experimental data on quasars that span 23\% to 87\% of the age of
the Universe, finding a variation in the fine structure constant, given by $%
\Delta \alpha /\alpha \cong -0.72$x10$^{-5}.$ We shall show that this result
is coherent with other cosmological facts.

One can ask wether this variation could be caused by a variyng speed of
light. Gomide [18] has long ago studied cosmological models with varying $c$%
, and/or varying $\varepsilon _{0}.$ Dirac's LNH (Large Numbers
Hypothesis)[9][10][11] was included in his framework. Barrow[6] and Barrow
and Magueijo [19] have studied how a time varying c would explain both, the $%
\dot{\alpha}\neq 0$ and the Supernovae observations. We refer to their
papers for further information. Our framework will presently be different
from those references.

If we perceive the present Universe as having constant deceleration
parameter, $q,$ like, for instance, Einstein-de Sitter's Universe, where $%
q=1/2$=$const.$, we may use Berman's\cite{3} \cite{4}formula \ for the
Hubble's parameter,

\begin{equation}
H=\frac{1}{mt}=\frac{1}{1+q}t^{-1}
\end{equation}
where

\begin{equation}
H=\dot R/R
\end{equation}
and

\begin{equation}
q=-\frac{\ddot RR}{\dot R^{2}}
\end{equation}
Overdots are for time-derivatives, and $R$ is the scale-factor in Robertson
Walker's metric,

\begin{equation}
ds^{2}=dt^{2}-\frac{R^{2}(t)}{(1+\frac{kr^{2}}{4})^{2}}(dx^{2}+dy^{2}+dz^{2})
\end{equation}
where $k=0,\pm 1$ is the tricurvature.

We may now express Webb's et al's\cite{2} experimental result as:

\begin{equation}
\left( \frac{\dot\alpha }{\alpha }\right) _{\exp }\simeq
-1.1\times10^{-5}H(1+q)
\end{equation}
We define

\begin{equation}
\alpha \equiv \frac{e^{2}}{\hbar c}
\end{equation}
and we shall suppose that the time-variation of $\alpha $ is caused by a
varying speed of light, as in Barrow's paper's\cite{5}\cite{6}\cite{7}.

From (6) we find,

\begin{equation}
\frac{\dot{\alpha}}{\alpha }=-\frac{\dot{c}}{c}.
\end{equation}
If we suppose that the speed of light varies with a power-law of time, say:

\begin{equation}
c=At^{n}
\end{equation}
($A=const.$); we find, from the above experimental values,

\begin{equation}
n=1.1\times 10^{-5}.
\end{equation}
We see that the speed of light varies slowly with the age of the Universe,
as Berman and Trevisan \cite{8} have shown elsewhere, for another model.

Now, let us consider Dirac's large number's hypothesis\cite{9}\cite{10}\cite
{11}. The number of nucleons in the Universe is called $N,$ and we find
roughly that both the ratio of the Hubble's length and the classical
electronic radius and the ratio of electrostatic and gravitational
interactions between a proton and a electron are of order $\sqrt{N}$ , with $%
N\sim 10^{80},$i. e.

\begin{equation}
\frac{cH^{-1}}{\frac{e^{2}}{m_{e}c^{2}}}\thickapprox \sqrt{N}
\end{equation}

\begin{equation}
\frac{e^{2}}{Gm_{p}m_{e}}\thickapprox \sqrt{N}
\end{equation}

\begin{equation}
\frac{\rho (cH^{-1})^{3}}{m_{p}}\thickapprox N
\end{equation}
Dirac's LNH can be presented by the assumption that $N$ varies with the age
of the Universe, so that a present ``large '' number N only means that the
Universe is ``old ''.

When we plug law (1) into (10)-(11)-(12), we find that,

\begin{equation}
\rho \varpropto t^{-2}\sqrt{N}
\end{equation}

\begin{equation}
G\varpropto (N)^{-1}
\end{equation}

\begin{equation}
c\varpropto (N)^{\frac{1}{6}}t^{-\frac{1}{3}}
\end{equation}
In order to accommodate (15) with (8) and (9), we assume that

\begin{equation}
N\varpropto t^{2.0001}
\end{equation}
and, then,

\begin{equation}
G\varpropto t^{-1.00005}
\end{equation}

\begin{equation}
\rho \varpropto t^{-0.99995}
\end{equation}
This ``solves '' \ \ Dirac's LNH, for the experimental found fine-structure
``constant 's '' time-variation. We find $\dot{G}/G\cong 1.0t^{-1}$ $\cong
1.0H(1+q)$ while it has been found under Lunar laser ranging and Viking
radar measurements by Williams et al. [12] and Reasenberg [13], that $\dot{G}%
/G=\sigma H$ with \TEXTsymbol{\vert}$\sigma |<0.6.$ Will [14][15] comments
that these two kinds of measurements give the best limits on $\sigma \lbrack
20].$ This means that -- 0.4\TEXTsymbol{>}$q>-1.6.$ A negative $q$ is a good
result because of Supernovae observations [16]. We have thus shown that $%
\Delta \alpha /\alpha $ should really be negative, for a positive result
could mean a positive deceleration parameter. As a bonus, we found how $N$, $%
G,\rho ,$ and the speed of light may vary with the age of the Universe, in
order to be in accordance with LNH and Webb et al's results.

In the model presented here, $\alpha =\alpha (t)$, because $c=c(t).$ It is
important to remark that the electric permittivity of the vacuum, as well as
its magnetic permeability, and Planck's constant, are thought really
constants; other possibility was devised by Berman and Trevisan [17]. It is
important to elaborate different models and make some predictions in order
to decide among them. In fact a Superunification theory will only survive in
case that such evolution of the constants values with the age of the
Universe will be explained by this theory.

\textbf{Acknowledgments}

Both authors thank support by Prof. Ramiro Wahrhaftig, Secretary of Science,
Technology , and Higher Education of the State of Paran\'{a}, and by our
Institutions, especially to Jorge\ L.Valgas, Roberto Merhy, Mauro K.
Nagashima, Carlos Fior, C.R. Kloss, J.L.Buso, and Roberto Almeida.

\end{document}